\documentclass[fleqn,usenatbib]{mnras}

\usepackage{newtxtext,newtxmath}

\usepackage[T1]{fontenc}

\DeclareRobustCommand{\VAN}[3]{#2}
\let\VANthebibliography\thebibliography
\def\thebibliography{\DeclareRobustCommand{\VAN}[3]{##3}\VANthebibliography}

\usepackage{graphicx}	
\usepackage{amsmath}	

\usepackage{orcidlink}

\usepackage{xcolor}
\usepackage{ulem}
\usepackage{mathrsfs}

\makeatletter
\DeclareRobustCommand{\HI}{%
  \mbox{H\check@mathfonts\fontsize\sf@size\z@\selectfont I}%
}
\makeatother
\newcommand{\lya}{Ly$\alpha$}
\newcommand{\lyb}{Ly$\beta$}

\newcommand{\kms}{km~s$^{-1}$}

\newcommand{\rom}[1]{\uppercase\expandafter{\romannumeral #1\relax}}

\def\cmpch{{h^{-1}{\rm Mpc}}} %
\def\teff{\tau_{\rm eff} }

\defcitealias{becker_mean_2021}{B21}
\defcitealias{malloy_how_2015}{ML15}

\title[Damping Wing-Like Features at $z<6$]{Damping Wing-Like Features in the Stacked \lya\ Forest: Potential Neutral Hydrogen Islands at $z<6$}

\author[Y.~Zhu et al.]{
Yongda Zhu\orcidlink{0000-0003-3307-7525},$^{1,2}$\thanks{E-mail: yongdaz@arizona.edu}
George D. Becker\orcidlink{0000-0003-2344-263X},$^{2}$
Sarah E. I. Bosman\orcidlink{0000-0001-8582-7012},$^{3,4}$
Christopher Cain\orcidlink{0000-0001-9420-7384},$^{5,2}$
Laura C. Keating\orcidlink{0000-0001-5211-1958},$^{6}$
\newauthor
Fahad Nasir\orcidlink{0000-0003-0294-8674},$^{3}$
Valentina D'Odorico\orcidlink{0000-0003-3693-3091},$^{7,8,9}$
Eduardo Ba\~{n}ados\orcidlink{0000-0002-2931-7824},$^{3}$
Fuyan Bian\orcidlink{0000-0002-1620-0897},$^{10}$
Manuela Bischetti\orcidlink{0000-0002-4314-021X},$^{11}$
\newauthor
James S. Bolton\orcidlink{0000-0003-2764-8248},$^{12}$
Huanqing Chen\orcidlink{0000-0002-3211-9642},$^{13}$
Anson D'Aloisio\orcidlink{0000-0003-2344-263X},$^{2}$
Frederick B. Davies\orcidlink{0000-0003-0821-3644},$^{3}$
\newauthor
Rebecca L. Davies\orcidlink{0000-0002-3324-4824},$^{14,15}$
Anna-Christina Eilers\orcidlink{0000-0003-2895-6218},$^{16}$
Xiaohui Fan\orcidlink{0000-0003-3310-0131},$^{1}$
Prakash Gaikwad\orcidlink{0000-0002-2423-7905},$^{3}$
Bradley Greig\orcidlink{0000-0002-4085-2094},$^{17,15}$
\newauthor
Martin G. Haehnelt\orcidlink{0000-0001-8443-2393},$^{18}$
Girish Kulkarni\orcidlink{0000-0001-5829-4716},$^{19}$
Samuel Lai\orcidlink{0000-0001-9372-4611},$^{20}$
Ewald Puchwein\orcidlink{0000-0001-8778-7587},$^{21}$
Yuxiang Qin\orcidlink{0000-0002-4314-1810},$^{17,15}$
\newauthor
Emma V. Ryan-Weber\orcidlink{0000-0002-5360-8103},$^{14,15}$
Sindhu Satyavolu\orcidlink{0000-0001-5818-6838},$^{19}$
Benedetta Spina\orcidlink{0000-0003-1634-1283},$^{4}$
Fabian Walter\orcidlink{0000-0003-4793-7880},$^{3}$
\newauthor
Feige Wang\orcidlink{0000-0002-7633-431X},$^{1}$
Molly Wolfson\orcidlink{0000-0003-0766-2499},$^{22}$
and Jinyi Yang\orcidlink{0000-0001-5287-4242}$^{1}$
\\
$^{1}$Steward Observatory, University of Arizona, 933 North Cherry Avenue, Tucson, AZ 85721, USA\\
$^{2}$Department of Physics \& Astronomy, University of California, Riverside, CA 92521, USA\\
$^{3}$Max-Planck-Institut für Astronomie, Königstuhl 17, D-69117 Heidelberg, Germany\\
$^{4}$Institute for Theoretical Physics, Heidelberg University, Philosophenweg 12, D-69120, Heidelberg, Germany\\
$^{5}$School of Earth and Space Exploration, Arizona State University, Tempe, AZ 85281, USA\\
$^{6}$Institute for Astronomy, University of Edinburgh, Blackford Hill, Edinburgh, EH9 3HJ, UK\\
$^{7}$INAF-Osservatorio Astronomico di Trieste, Via Tiepolo 11, I-34143 Trieste, Italy\\
$^{8}$Scuola Normale Superiore, Piazza dei Cavalieri 7, I-56126 Pisa, Italy\\
$^{9}$IFPU-Institute for Fundamental Physics of the Universe, via Beirut 2, I-34151 Trieste, Italy\\
$^{10}$European Southern Observatory, Alonso de Córdova 3107, Casilla 19001, Vitacura, Santiago 19, Chile\\
$^{11}$Dipartimento di Fisica, Sezione di Astronomia, Università di Trieste, via Tiepolo 11, 34143 Trieste, Italy\\
$^{12}$School of Physics and Astronomy, University of Nottingham, University Park, Nottingham, NG7 2RD, UK\\
$^{13}$Canadian Institute for Theoretical Astrophysics, University of Toronto, Toronto, ON M5R 2M8, Canada\\
$^{14}$Centre for Astrophysics and Supercomputing, Swinburne University of Technology, Hawthorn, Victoria 3122, Australia\\
$^{15}$ARC Centre of Excellence for All Sky Astrophysics in 3 Dimensions (ASTRO 3D), Australia\\
$^{16}$MIT Kavli Institute for Astrophysics and Space Research, 77 Massachusetts Avenue, Cambridge, MA 02139, USA\\
$^{17}$School of Physics, University of Melbourne, Parkville, VIC 3010, Australia\\
$^{18}$Kavli Institute for Cosmology and Institute of Astronomy, Madingley Road, Cambridge, CB3 0HA, UK\\
$^{19}$Tata Institute of Fundamental Research, Homi Bhabha Road, Mumbai 400005, India\\
$^{20}$Commonwealth Scientific and Industrial Research Organisation (CSIRO), Space \& Astronomy, P. O. Box 1130, Bentley, WA 6102, Australia\\
$^{21}$Leibniz-Institut für Astrophysik Potsdam, An der Sternwarte 16, 14482 Potsdam, Germany\\
$^{22}$Department of Physics, University of California, Santa Barbara, CA 93106, USA
}

\date{Accepted XXX. Received YYY; in original form ZZZ}

\pubyear{2024}

\begin{document}
\label{firstpage}
\pagerange{\pageref{firstpage}--\pageref{lastpage}}
\maketitle

\begin{abstract}
Recent quasar absorption line observations suggest that reionization may end as late as $z \approx 5.3$. As a means to search for large neutral hydrogen islands at $z<6$, we revisit long dark gaps in the Ly$\beta$ forest in VLT/X-Shooter and Keck/ESI quasar spectra. We stack the Ly$\alpha$ forest corresponding to both edges of these \lyb\ dark gaps and identify a damping wing-like extended absorption profile. The average redshift of the stacked forest is $z=5.8$. By comparing these observations with reionization simulations, we infer that such a damping wing-like feature can be naturally explained if these gaps are at least partially created by neutral islands.
Conversely, simulated dark gaps lacking neutral hydrogen struggle to replicate the observed damping wing features. Furthermore, this damping wing-like profile implies that the volume-averaged neutral hydrogen fraction must be $\langle x_{\rm HI} \rangle \geq 6.1 \pm 3.9\%$ at $z = 5.8$. Our results offer robust evidence that reionization extends below $z=6$.
\end{abstract}




\section{Introduction \label{sec:introduction_p4}}

Hydrogen reionization carries key implications for the formation and evolution of the first stars, galaxies, and supermassive black holes (see, e.g., \citealp{fan_quasars_2023} for a review). Cosmic microwave background (CMB) observations indicate a midpoint of reionization at $z \sim 7-8$ \citep[][see also \citealp{de_belsunce_inference_2021}]{planck_collaboration_planck_2020}. 
Meanwhile, transmitted flux observed in the \lya\ forest at $z \sim 6$ towards high-redshift quasars \citep[e.g.,][]{fan_constraining_2006} has long been interpreted as an indicator of the end of reionization. 
However, if reionization ended by $z\sim 6$, galaxies would have had to produce an extremely large amount of ionizing photons to complete reionization within a short timeframe \citep[see also][]{munoz_reionization_2024-1}. This may pose a significant challenge to our understanding of star formation, the escape fraction of ionizing photons, and the spectral energy distribution (SED) of early galaxies \citep[e.g.,][]{robertson_cosmic_2015,bouwens_reionization_2015,finkelstein_conditions_2019,stark_galaxies_2016}.

A later end to reionization, which would help resolve such tension, is gaining support from recent observations. Supporting evidence includes large-scale fluctuations in the 
\lya\ effective optical depth\footnote{Defined as $\teff=-\ln{\langle F \rangle}$, where $F$ is the continuum-normalized transmission flux.} 
measured in quasar spectra 
\citep[e.g.,][]{fan_constraining_2006,becker_evidence_2015,eilers_opacity_2018,bosman_new_2018,bosman_hydrogen_2022,yang_measurements_2020}; long troughs extending down to or below $z\simeq5.5$ in the \lya\ and \lyb\ forests \citep[e.g.,][]{becker_evidence_2015,zhu_chasing_2021,zhu_long_2022}, potentially indicating the existence of large neutral intergalactic medium (IGM) islands \citep[e.g.,][]{kulkarni_large_2019,keating_long_2020,nasir_observing_2020,qin_reionization_2021}; 
observed underdensities around long dark gaps traced by \lya\ emitting galaxies \citep[LAEs,][]{becker_evidence_2018,kashino_evidence_2020, christenson_constraints_2021,christenson_relationship_2023-1}; the evolution of metal-enriched absorbers at $z\sim6$ \citep[e.g.,][]{becker_evolution_2019,cooper_heavy_2019,davies_xqr-30_2023,davies_examining_2023-1,sebastian_e-xqr-30_2024}; and the dramatic evolution in the measured mean free path of ionizing photons over $5<z<6$ \citep[][see also \citealp{bosman_constraints_2021-1,gaikwad_measuring_2023,satyavolu_robustness_2023,roth_effect_2024,davies_constraints_2024}]{becker_mean_2021,zhu_probing_2023}, which provides the most unambiguous evidence to date of ongoing reionization at $z<6$. A late-end reionization scenario is also consistent with numerical models that reproduce a variety of observations \citep[e.g.,][Y.~Qin et al.~in prep.]{weinberger_modelling_2019,choudhury_studying_2021,qin_reionization_2021,gaikwad_measuring_2023,asthana_late-end_2024}.

A pressing question is whether we can detect neutral hydrogen (\HI) islands at $z<6$ if they exist. In terms of resonant \lya\ and \lyb\ absorption, the observational signatures of neutral islands may be difficult to distinguish from those of large-scale fluctuations in the ionizing ultraviolet background; either one may produce extended regions of nearly zero transmission. At higher redshifts, where the IGM is more neutral, damping wing features over the \lya\ emission of quasars and galaxies have been observed and used as a powerful probe of the volume-averaged neutral hydrogen fraction ($x_{\rm HI}$) of the Universe \citep[e.g.,][]{banados_800-million-solar-mass_2018,davies_quantitative_2018,wang_significantly_2020,yang_poniuaena_2020,greig_igm_2022,greig_igm_2024,umeda_jwst_2023,durovcikova_chronicling_2024-1}. Any signature of
damping wing absorption at lower redshifts might similarly indicate the presence of neutral islands.  A detection of damping wings would also enable constraints of $x_{\rm HI}$ that would complement the upper limits from the fraction of dark gaps and dark pixels in the forest \citep[][]{mcgreer_z_2013,zhu_long_2022,jin_nearly_2023}.

\citet[][hereafter referred to as \citetalias{malloy_how_2015}]{malloy_how_2015} proposed to test for neutral islands at $z\sim 5.5$ by searching for damping wing absorption over the stacked \lya\ forest adjacent to highly absorbed regions. This method has not been implemented in observations because it requires a large sample of high-quality spectra of high-$z$ quasars and a relatively complete catalog of intervening metal absorbers to exclude contamination from damped \lya\ systems (DLAs) arising from galaxies. Thanks to a large sample of quasar spectra provided in the (Extended) XQR-30 large program \citep{dodorico_xqr-30_2023} and the metal absorber catalog of \citet{davies_xqr-30_2023}, we are now able to carry out the experiment proposed in \citetalias{malloy_how_2015}. Long dark gaps have been detected in the \lyb\ forest \citep{zhu_long_2022}, and they indicate regions of high IGM opacity that may potentially host neutral islands \citep[e.g.,][]{kulkarni_large_2019,keating_long_2020,nasir_observing_2020,qin_reionization_2021}. Therefore, stacking the \lya\ forest at the redshifts of these long dark gaps is a powerful way to implement the \citetalias{malloy_how_2015} test and search for IGM damping wing features at $z<6$. 
Instead of only using the red damping wings (e.g., like in quasar damping wing measurements), we include both ends of the dark gaps, because the damping wing signal from a finite-sized pocket of neutral gas can be observed on both sides of the dark gap.

This letter is organized as follows. In {Section~\ref{sec:data_p4}} we describe the data and \lyb\ dark gaps used in this work. \mbox{Section~\ref{sec:results_p4}} presents the stacked \lya\ profile. Section~\ref{sec:comparison} compares our results to model predictions, and Section~\ref{sec:lower_limit} presents our constraints on $x_{\rm HI}$. Finally, we conclude our findings in \mbox{Section~\ref{sec:summary_p4}}. Throughout this paper, we quote distances in comoving units unless otherwise noted and assume a $\Lambda$CDM cosmology with $\Omega_{\rm m}=0.308$, $\Omega_{\Lambda}=0.692$, and $h=0.678$ \citep[][]{planck_collaboration_planck_2014}. In addition to the stacking method used in this work, our companion paper, \citet{becker_damping_2024} present the discovery of an IGM damping wing at $z<6$ towards the individual quasar sightline, J0148+0600.

\section{Data}\label{sec:data_p4}

To create a stacked spectrum of the \lya\ forest over potentially neutral regions at $z<6$, we revisit dark gaps detected in the \lyb\ forest in \citet{zhu_long_2022}. 
Compared to \lya\ absorption, the lower optical depth to \lyb\ photons increases the likelihood that \lyb\ dark gaps are caused by high-opacity neutral pockets rather than ionized regions with low ionizing background.
We refer the readers to \citet{zhu_chasing_2021,zhu_long_2022} for details of the data and dark gap detection. Briefly, \lyb\ dark gaps are detected in 42 spectra of quasars at $5.77 \lesssim z_{\rm quasar} \lesssim 6.31$. The spectra are taken with the Echellette Spectrograph and Imager (ESI) on Keck \citep{sheinis_esi_2002} and the X-Shooter spectrograph on the Very Large Telescope (VLT; \citealp{vernet_x-shooter_2011}). Among these, 19 X-Shooter spectra are from the XQR-30 large program \citep{dodorico_xqr-30_2023}. A dark gap in the \lyb\ forest is defined as a continuous spectral region in which all pixels binned to $1 \cmpch$ have an observed normalized flux $F = F_{\rm obs}/F_{\rm c} < 0.02$, where $F_{\rm obs}$ is the observed flux and $F_{\rm c}$ is the continuum flux predicted from Principal Component Analysis (PCA; see \citealp{bosman_comparison_2021,bosman_hydrogen_2022,zhu_chasing_2021}). The \lya\ forest at the corresponding redshift of a \lyb\ dark gap also needs to be opaque ($F_{\rm Ly\alpha} = F_{\rm obs}/F_{\rm c} < 0.05$). 

In this work we use only the \citet{zhu_long_2022} dark gaps at $z < 6$.
Dark gap statistics based on mock spectra generated from simulations suggest that longer dark gaps are more likely to have a higher covering fraction of neutral hydrogen \citep{nasir_observing_2020,zhu_long_2022}. The stack based on longer dark gaps might also create a stronger damping wing profile in the corresponding \lya\ forest \citepalias{malloy_how_2015}. To balance the sample size and dark gap length, here we use \lyb\ dark gaps with $L \geq 7 \cmpch$. Considering the low transmission in the \lya\ forest at these redshifts, we only include quasar spectra with a continuum-normalized flux error $<0.05$ per pixel in the \lya\ and \lyb\ forest. Therefore, dark gaps toward quasars PSO J308-21 and VIK J2318-3029 are excluded due to the relatively low S/N of the spectra. 

We have carefully checked these dark gaps to ensure they do not contain intervening metal absorbers within 3000 \kms\ near the edge of dark gaps, based on the metal line catalog presented in \citet{davies_xqr-30_2023} and through visual inspection. This requirement rejects the dark gap toward quasar SDSS J0818+1722 spanning $z=5.761-5.794$ with $L=10\cmpch$. This dark gap contains absorption lines of \ion{O}{I}, \ion{C}{II}, \ion{Si}{II}, \ion{C}{IV}, etc., near the red edge at $z \simeq 5.79$. 
We note that, however, including this dark gap produces no substantial difference in the results. A summary of \lyb\ dark gaps used in this work is provided in Table \ref{tab:gaplist}.  Our final sample includes 24 dark gaps, for which the average redshift at the red and blue edges is $z=5.8$.

\begin{table}
    \centering
    \caption{Ly$\beta$ Dark Gaps Used in this Work}
    \label{tab:gaplist}
    \begin{tabular}{clcccr}
        \hline
        No. & Quasar & $z_{\rm q}^{\rm ref}$ & $z_{\rm blue}$ & $z_{\rm red}$ & $L_{\rm gap}$ \\
        \hline
 1 & ULASJ1319+0950 & $6.1330^{\rm vi}$ &5.876 & 5.903 & 8  \\
   2 & PSOJ060+24 & $6.1793^{\rm ii}$ &$5.833^{\rm a}$ & 5.856 & 7  \\
   3 & PSOJ108+08 & $5.9647^{\rm ii}$ &5.674 & 5.751 & 24 \\
   4 & SDSSJ0842+1218 & $6.0763^{\rm iii}$ &5.784 & 5.830 & 14 \\
   5 & SDSSJ2315-0023 & $6.124^{\rm i}$ &5.790 & 5.883 & 28 \\
   6 & SDSSJ2315-0023 & $6.124^{\rm i}$ &5.897 & $5.937^{\rm b}$ & $\geq12 $\\
   7 & CFHQSJ1509-1749 & $6.1225^{\rm iii}$ &5.800 & 5.863 & 19 \\
   8 & CFHQSJ1509-1749 & $6.1225^{\rm iii}$ &5.870 & 5.910 & 12 \\
   9 & SDSSJ2054-0005 & $6.0391^{\rm vi}$ &5.751 & 5.790 & 12 \\
  10 & SDSSJ0840+5624 & $5.8441^{\rm vi}$ &5.585 & 5.607 & 7  \\
  11 & PSOJ340-18 & $5.999^{\rm i}$ &5.774 & 5.810 & 11 \\
  12 & PSOJ065-26 & $6.1877^{\rm iii}$ &5.954 & 5.988 & 10 \\
  13 & PSOJ007+04 & $6.0008^{\rm iii}$ &5.741 & 5.780 & 12 \\
  14 & ULASJ0148+0600 & $5.9896^{\rm ii}$ &5.654 & 5.735 & 25 \\
  15 & ULASJ0148+0600 & $5.9896^{\rm ii}$ &5.741 & 5.803 & 19 \\
  16 & PSOJ217-16 & $6.1498^{\rm iii}$ &5.807 & 5.873 & 20 \\
  17 & PSOJ217-16 & $6.1498^{\rm iii}$ &5.920 & $5.961^{\rm b}$ & $\geq12 $\\
  18 & J0408-5632 & $6.0264^{\rm ii}$ &5.715 & 5.741 & 8  \\
  19 & PSOJ359-06 & $6.1718^{\rm iv}$ &5.886 & 5.917 & 9  \\
  20 & PSOJ025-11 & $5.8414^{\rm ii}$ &5.526 & 5.613 & 28 \\
  21 & PSOJ025-11 & $5.8414^{\rm ii}$ &5.632 & 5.661 & 9  \\
  22 & PSOJ158-14 & $6.0681 ^{\rm iv}$ &5.764 & 5.836 & 22 \\
  23 & PSOJ158-14 & $6.0681 ^{\rm iv}$ &5.843 & 5.880 & 11 \\
  24 & SDSSJ1250+3130 & $6.137^{\rm v}$ &5.836 & 5.870 & 10 \\
        \hline
    \end{tabular}
    
    \textit{Notes.} Columns: 
    (1) Index of dark gaps; 
    (2) quasars used in \citet{zhu_chasing_2021}; 
    (3) quasar redshift with reference;
    (4) redshift at the blue end of the gap; 
    (5) redshift at the red end of the gap; 
    (6) dark gap length in units of $h^{-1}\,\mathrm{Mpc}$. $^{\rm a}$ Dark gap starting at the blue edge of the Ly$\beta$ forest. $^{\rm b}$ Dark gap ending at the red edge of the Ly$\beta$ forest as defined in \citet{zhu_long_2022}. 
    Quasar redshift references:
    i. \citet{becker_evolution_2019}, 
    ii. S.~Bosman et al.~(in prep),
    iii. \citet{decarli_alma_2018},
    iv. \citet{eilers_detecting_2020},
    v. \citet{shen_gemini_2019},
    vi. \citet{wang_molecular_2010,wang_star_2013}.
\end{table}

\begin{figure}
    \hspace{-0.5cm}
    \includegraphics[width=3.4in]{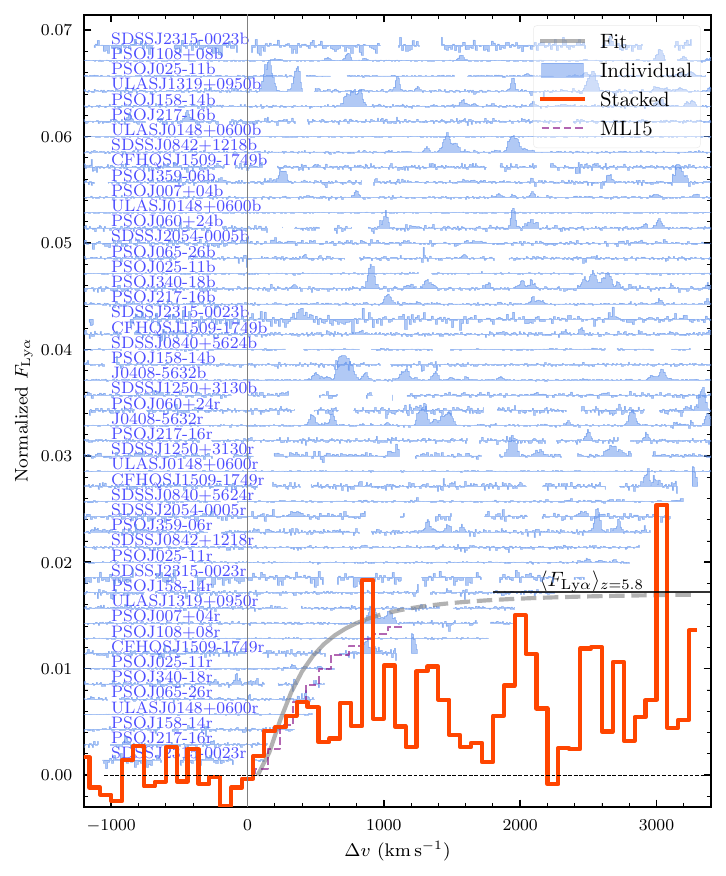}
    \vspace{-0.5cm}
    \caption{Stacked \lya\ forest corresponding to the redshift of both edges of \lyb\ dark gaps (displayed in red). The grey curve represents the fitted damping wing profile, as described by Equation \ref{eq:damping_wing}, applied to the upper envelope of the stacked flux between $\Delta v=0$ \kms\ and $1000$ \kms.
    The individual spectra used in this stack are shown in blue (sorted by their $\Delta v$ coverage), with the flux scaled by a factor of 0.005 for display purposes. We have masked regions that may suffer from sky line subtraction residuals. The letters ``r'' and ``b'' following the quasar name indicate that we are stacking the \lya\ forest at the redshift corresponding to the red and blue edges of the \lyb\ gap, respectively. The stacked \lya\ transmission profile is comparable with that expected for neutral islands with $L=5.34h^{-1}\rm Mpc$ (dashed purple line) in \citetalias{malloy_how_2015}.}
    \label{fig:spectra}
\end{figure}

\section{Stacked Ly\texorpdfstring{$\alpha$}{Lg} Forest}\label{sec:results_p4}

Following \citetalias{malloy_how_2015}, we create a stacked \lya\ transmission profile by aligning the spectra based on the redshift at the edges of \lyb\ dark gaps.\footnote{When aligning the spectra at the redshift corresponding to the blue edge of \lyb\ dark gaps, we flip the sign of the velocity offset ($\Delta v$) such that the velocity offset increases when moving away from the dark gap edges.} 
We firstly bin the continuum-normalized \lya\ forest to 80 \kms\ intervals, corresponding to $\sim 0.5\cmpch$ at $z=5.8$, which is half of the bin size used by \citet{zhu_long_2022} when searching for dark gaps. To remove the redshift evolution in the mean \lya\ transmission, we re-normalize the sightlines with a factor equal to the mean transmission at $z=5.8$ divided by the mean transmission at the redshift of each pixel according to the measurements in \citet{bosman_hydrogen_2022} with a linear interpolation. The normalization factor is small and close to 1, ensuring minimal distortion of the data. The 
stacked spectrum is then created by taking a mean value of the flux in each bin. We exclude the quasar proximity zone effect by conservatively disregarding the \lya\ forest within 11 proper-Mpc towards the quasar following \citet{zhu_long_2022}. We also mask pixels that are affected by sky-line subtraction residuals or telluric absorption correction as indicated by peaks in the flux error array.

Figure \ref{fig:spectra} shows the stacking result along with the \lya\ forest corresponding to each individual \lyb\ dark gap. We see an extended damping wing-like absorption feature redward of the edge of the dark gaps, with the transmission gradually recovering 
to the mean value at a velocity offset of $\Delta v \sim 1000$ \kms. Notably, such a damping wing-like feature is not obvious in any of the individual spectra.\footnote{Although strong transmission spikes appear adjacent to the blue edge  of the \lyb\ dark gap toward ULAS J1319+0950, we have tested that excluding this sightline will not change the damping wing-like profile significantly.}

Following \citetalias{malloy_how_2015} (see also \citealp{miralda-escude_searching_1998,miralda-escude_reionization_1998}), the absorption due to the presence of a neutral hydrogen island will be extended, with optical depth far from the line center described by a damping wing. The optical depth at a velocity offset $\Delta v$ from the edge of a neutral island will be  
\begin{equation}
    \tau_{\rm Ly\alpha}^{\rm DW} (\Delta v) \approx
    \frac{\tau_{\rm GP} R_\alpha c }
    {\pi}
    \left[ \frac{1}{\Delta v} - \frac{1}{\Delta v + v_{\rm ext}}\right]\,,
    \label{eq:damping_wing}
\end{equation}
where $\tau_{\rm GP}$ is the Gunn-Peterson optical depth,
for which we take a nominal value of $2.5\times 10^5$ at $z \sim 5.8$, $R_\alpha \equiv \Gamma_\alpha \lambda_\alpha / 4\pi c$, and $\Gamma_\alpha = 6.265 \times 10^8$ s$^{-1}$ is the \lya\ decay constant. 
In contrast to \citetalias{malloy_how_2015}, where $v_{\rm ext}$ measures the extent of neutral islands in the velocity space, we adopt $v_{\rm ext}$ as an \textit{empirical parameter} to describe the extent of the transmission profile. 

This change is mainly due to a different fitting method. Instead of fitting the mean transmission, we fit the damping wing profile in Equation \ref{eq:damping_wing} to the \textit{upper envelope} of the stacked spectrum in the interval between $\Delta v = 0$ \kms\ and $1000$ \kms. 
We find this approach more appropriate for our data because the stacked profile is noisy at a large $\Delta v$, making the mean transmission fitting less reliable. Also, the damping wing profile cannot be significantly lower than any transmission peaks in real data.
Specifically, we perform a least-squares fitting over the stacked flux, and require that the fitted curve should not be significantly lower than the stacked flux in any pixel \citep[e.g.,][]{prochaska_esikeck_2003}. 
Here, we set the threshold to be the smaller of $80\%$ of the stacked flux or $1-\sigma$ white noise below the flux. The white noise is given by the standard deviation measured over the transmission-free region in the stacked \lya\ forest over $\Delta v = -1000$ \kms\ to $0$ \kms. We have tested that using a different threshold does not change our conclusions. As Figure \ref{fig:spectra} shows, the fitted curve well describes the observed extended absorption feature and yields $v_{\rm ext} = 357$ \kms. We note that the stacked \lya\ transmission profile is comparable with that expected for neutral islands with $L=5.34\cmpch$ in \citetalias{malloy_how_2015}, considering that the average span of our \lyb\ dark gaps and the required \HI\ covering fraction yield a length of neutral island to be around $5\cmpch$ (see Section \ref{sec:lower_limit}), although our profile is a bit steeper considering the difference in the fitting methods.
Similar damping wing-like signals at $z\lesssim6$ are also presented in \citet{spina_damping_2024}, although a different method is used in their paper.

We emphasize that our approach does not represent a physical fit to a damping wing per se, but rather a fit to the transmission profile
using a \textit{damping wing-like} function. The observed mean profile potentially includes damping wing absorption arising from genuine neutral islands,
but will also include resonant \lya\ absorption, which may be modified on the scales of interest by UVB fluctuations. As described in Section \ref{sec:comparison}, we assess the evidence for damping wing absorption by fitting the same function to mock stacks drawn from simulations.

We note that there are large fluctuations in the stacked transmission profile at $\Delta v \gtrsim 1000~{\rm km\,s^{-1}}$. These are likely statistical fluctuation resulting from 
the relatively small number of spectra included in the stack. Alternatively, these absorption dips may come from individual strong absorbers in the ionized regions redward of the gaps in the stack. A clustering of neutral islands on the scales of $\sim 10\cmpch$, however, may also create absorption features redward the damping wing-like profile in the stack, although this hypothesis would require more data to test.

\section{Comparison with Simulations}\label{sec:comparison}

A critical question is whether 
the observed damping wing-like profile in the stacked \lya\ forest
indicates the presence of genuine neutral islands at $z<6$. To explore the conditions under which such a profile might arise, we generate mock spectra from three sets of simulations, create stacks based on the \lyb\ gaps in these spectra, and compare the properties of the mock stacks to the observations.

In our first model, the sightlines are drawn from a radiative transfer (RT) simulation conducted in a $(200h^{-1}\rm Mpc)^3$ volume box with $N=200^3$ RT cells, using the adaptive ray-tracing code described in \citet{cain_short_2021,cain_rise_2023-1,cain_morphology_2023}. Detailed methodology will be presented in C.~Cain et al.~(in prep.). In the simulation, halos and galactic sources are populated based on abundance matching according to the luminosity function from \citet{finkelstein_conditions_2019}, and the hydrogen ionizing emissivity 
of the sources is assumed to be proportional to their
UV luminosity. This simulation is calibrated to match the observed mean transmission in the \lya\ forest as reported in \citet{bosman_hydrogen_2022}. We account for the attenuation from the foreground \lya\ forest when computing the \lyb\ forest flux. Next, we adjust the modeled spectra to have the same redshift coverage as the observed sample and add noise according to the flux error array from the observations. 

Dark gaps in the simulated sightlines are identified following the same criteria as in \citet{zhu_long_2022}; 
specifically, we select gaps that are dark in both the \lyb\ forest and the \lya\ forest. We then stack the \lya\ forest at the edges of \lyb\ dark gaps with lengths $L\geq 7\cmpch$.
For our first model, we create stacked \lya\ forest by including dark gaps with a volume-weighted neutral hydrogen fraction of $\geq 0.7$ measured in the pencil beam cells (denoted as ``$x_{\rm HI,\,gap}\geq 0.7$''), and dark gaps that contains only ionized gas (denoted as ``Ionized dark gap'', respectively. In practice, we exclude dark gaps with any pencil beam cells that have volume-weighted neutral fraction greater than 0.5. The fraction of ``Ionized dark gap'' is 29\% out of all $L\geq 7\cmpch$ \lyb\ gaps, and the mean $x_{\rm HI,\,gap}$ is $2\times10^{-3}$.

The second simulation we use here employs a fluctuating UVB model, wherein the UVB fluctuations are boosted by a short mean free path and neutral islands are not explicitly included (denoted as ``ND20-early-shortmfp''; refer to \citealp{nasir_observing_2020,zhu_chasing_2021} for details). In contrast to the radiative transfer simulations of our first model, \citet{nasir_observing_2020} employ approximate ``semi-numeric'' methods to model reionization’s effects on the forest. For this model, we include all \lyb\ dark gaps with $L\geq 7\cmpch$, except for those with all pencil beam cells that have volume-averaged neutral fraction greater than 0.5. As a result, 99.88\% of all $L\geq 7\cmpch$ \lyb\ gaps are selected, and the mean $x_{\rm HI,\,gap}$ is $<0.03$.
Here, this model serves as an independent reference for dark gaps that are not dominated by neutral hydrogen islands.

In addition to the scenarios mentioned above, we include simulated sightlines from a simulation of inhomogeneous reionization from the Sherwood-Relics suite \citep[described in detail in][]{puchwein_sherwood-relics_2023} at different redshifts to investigate the effects of a varying overall neutral hydrogen fraction. The simulation we draw the sightlines from has a volume of 160 $h^{-1}$Mpc and was performed with 2048$^{3}$ particles. An inhomogeneous UV background is included in the simulation, taken from a pre-existing radiative transfer simulation performed on snapshots of a cosmological simulation that used the same initial conditions. This allows for spatial fluctuations in the ionization state of the gas due to patchy reionization and the associated hydrodynamic response of the gas to the photoheating. Reionization in the simulation completes at $z=5.3$. We draw mock spectra from $z=5.4$, 5.8, and 6.2, which have overall neutral fraction of $\langle x_{\rm HI} \rangle = 0.4\%$, 7.4\%, and 21.3\%, respectively. 

We generate mock datasets from the simulations to test the likelihood that a profile similar to the data will emerge in small samples. To mitigate the effect of variations in dark gap length distributions across different cases, in each bootstrap realization, we randomly draw dark gaps from the simulations, following exactly the same distribution of dark gap lengths as in the observed data to create the stack. Nevertheless, the shape of the stacked \lya\ forest does not change significantly even without the correction on the dark gaps length distribution.

\begin{figure}
    \centering
    \includegraphics[width=0.9\linewidth]{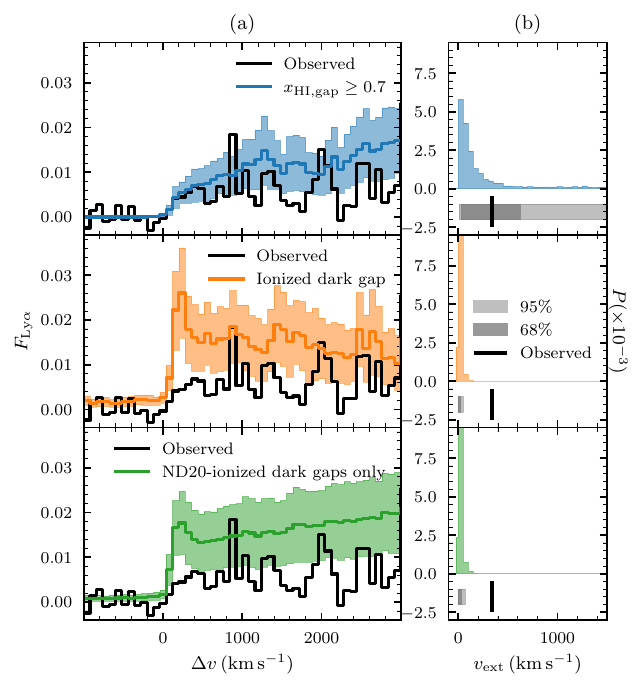}
    \vspace{-0.5cm}
    \caption{
    \textbf{(a)} Comparison of the stacked Ly$\alpha$ forest at the edges of Ly$\beta$ dark gaps in mock spectra to the observations. The shaded regions show the 68\% intervals of model predictions from bootstrap trials. 
    \textbf{(b)} Distributions of the damping wing parameter, $v_{\rm ext}$, in the models. The light and dark shaded horizontal bars in each subplot show the 68\% and 95\% intervals, respectively. The vertical black line denotes $v_{\rm ext}$ acquired from the observation.
    Based on these comparisons, we can see that simulations wherein dark gaps are substantially neutral are consistent with the observed damping wing-like profile.
    }
    \label{fig:cmp}
\end{figure}

In the Figure \ref{fig:cmp}, we compare the stacked \lya\ forest between observations and the mock stacks from a fixed redshift, i.e., $z=5.8$. As expected, 
the stack of dark gaps containing a significant fraction of neutral gas exhibits an extended and damping wing-like profile in the \lya\ forest. The observed curve broadly aligns with the model profile and falls within the 68\% interval. On the other hand, the ``ionized dark gap'' case shows a rapid increase in the stacked \lya\ transmission, inconsistent with the observed damping wing-like profile. The ``ND20-early-shortmfp'' case also displays a steep increase in the stacked \lya\ transmission, albeit with a slightly lower maximum flux compared to the ``ionized dark gap'' cases.

For a quantitative comparison, we fit the upper envelope of the model-predicted stacked \lya\ transmission in each bootstrap realization to Equation \ref{eq:damping_wing} and plot the distribution of the damping wing parameter $v_{\rm ext}$ in Figure \ref{fig:cmp}(b). Only the scenario where dark gaps contain a large fraction of neutral gas predicts a broad distribution of $v_{\rm ext}$, spanning the observed value within the 68\% interval. Conversely, scenarios that fail to reproduce an extended absorption profile predict a very narrow $v_{\rm ext}$ distribution, confined to $v_{\rm ext} \lesssim 100$ \kms, with the observed $v_{\rm ext}$ lying beyond their 95\% confidence limits. These results suggest that the observed damping wing-like profile in the stacked \lya\ transmission likely originates from genuine neutral IGM absorption at $z<6$, as opposed to fully ionized gas that is highly opaque to due to a fluctuating UVB or high densities. 

\begin{figure}
    \centering
    \includegraphics[width=0.9\linewidth]{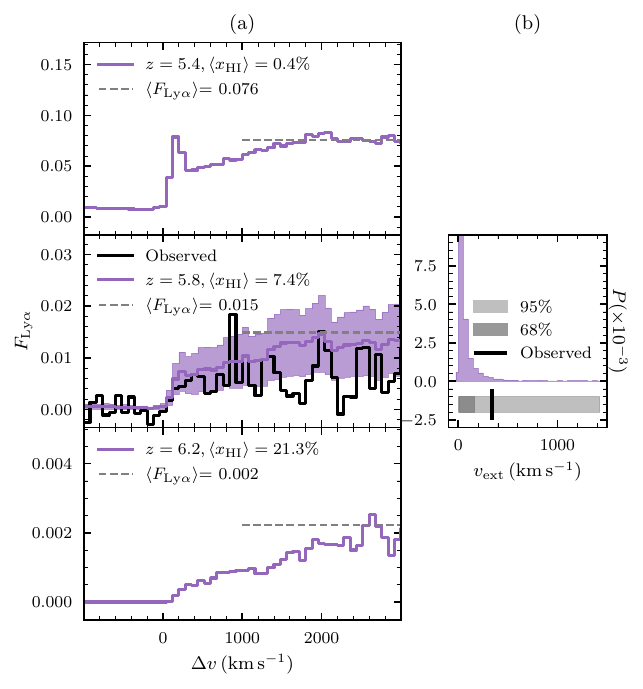}
    \vspace{-0.5cm}
    \caption{
    \textbf{(a)} Similar to Figure \ref{fig:cmp}, but comparing the observed stacked \lya\ forest with mock data that are drawn from simulations with different overall neutral fraction from the Sherwood-Relics patchy reionization simulation \citep{puchwein_sherwood-relics_2023}. Here, we include \emph{all} dark gaps regardless of their individual neutral hydrogen fraction. We display the shape of the mock stacked profiles from $z=5.4$ and 6.2 snapshots just for illustration purpose because their mean forest transmission and dark gap length distribution are significantly different from those at $z=5.8$, and different flux thresholds for dark gap detection are used at these redshifts.
    \textbf{(b)} $v_{\rm ext}$ distribution from the patchy reionization simulation at $z=5.8$. The observed damping wing-like profile is consistent with the model prediction that has $\langle x_{\rm HI} \rangle = 7.4\%$.
    }
    \label{fig:cmp2}
\end{figure}

As an additional test, we show the redshift evolution of this signal with mock stacks of spectra created from snapshots of one of the Sherwood-Relics simulations as described above. 
For this test we include all dark gaps present in the simulated lines of sight, without any pre-selection based on their individual neutral fractions. The results are shown in Figure~\ref{fig:cmp2}. 
As the volume-averaged ionization fraction of the IGM evolves, we observe changes in both the mean flux of the \lya\ forest and the shape of the \lya\ profile. For the snapshot at $z=5.4$, when the volume-averaged \HI\ fraction is 0.4\%, the \lya\ profile is consistent with the ``ionized dark gap'' scenario, where the \lya\ transmission is boosted on the edge of \lyb\ dark gaps. With increasing redshift and hence an increasing $\langle x_{\rm HI} \rangle$, the flux boost becomes smoother and finally approaches a damping wing-like shape. For the snapshot at $z=5.8$, which has a \lya\ forest mean flux consistent with the observational sample, the distribution of $v_{\rm ext}$ is consistent with the observation within the 95\% limits, possibly because the neutral fraction is just above the lower limit constraint (see Section \ref{sec:lower_limit}). This shows that simultaneously fitting for the mean flux of the \lya\ forest as well as the shape of the \lya\ profile at the edge of \lyb\ dark gaps can be a powerful probe of the end stages of reionization.

\begin{figure}
    \centering
    \includegraphics[width=0.38\textwidth]{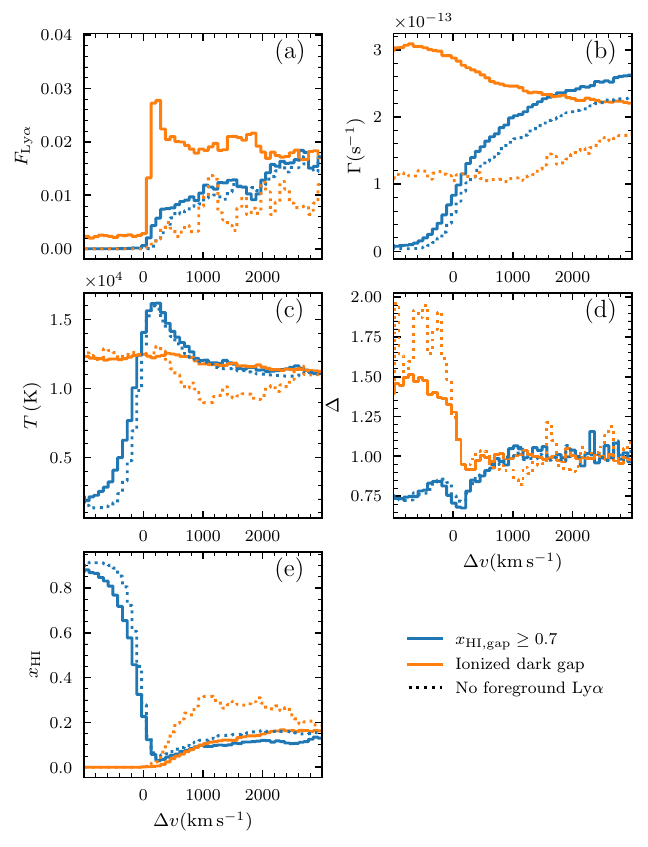}
    \vspace{-0.5cm}
    \caption{Stacked (mean) \lya\ transmission profiles at the redshift of the  edges of \lyb\ dark gaps detected in ideal sightlines from our first simulation. Panels show (a) the \lya\ transmission, (b) the ionization rate, (c) the IGM temperature, (d) the overdensity, and (e) the volume filling factor of neutral hydrogen. The solid blue line represents the ``$x_{\rm HI,\,gap}\geq0.7$'' case, where dark gaps are at least $70\%$ neutral. The solid orange line depicts the case of ``ionized dark gaps''. Dotted lines indicate the stacked profiles when \lyb\ dark gaps are selected without contamination from foreground \lya\ absorption.}
    \label{fig:insight}
\end{figure}

We further investigate the physical properties the simulated \lyb\ dark gaps by dividing the sample based on their neutral fraction.
For our first model, Figure \ref{fig:insight} displays the stacked properties of the \lyb\ dark gaps at the redshift of their edges. As the figure shows, dark gaps with substantial neutral hydrogen produce an extended absorption profile in the adjacent 
\lya\ forest. These dark gaps are typically found in regions characterized by a low ionizing UVB, low temperature, and low densities. Their properties align with findings in studies such as \citet{gnedin_cosmic_2022-1}. 

Notably, we find that if \lyb\ dark gaps are selected in the absence of contamination from foreground \lya\ absorption, then even the ``ionized dark gap'' scenario exhibits extended \lya\ absorption. Physically, this is because the \lya\ transmission should recover gradually when moving away from ionized regions of high \lya\ opacity. Contamination from \lyb\ gaps that are largely created by foreground \lya\ absorption obscures this effect. We find that up to $41\%$ of the total \lyb\ dark gap length in our first model can arise from the foreground contamination. As for the ``$x_{\rm HI,\,gap}\geq 0.7$'' scenario, the selection criteria ensure that the dark gaps contain genuine neutral islands that are not highly contaminated. As discussed above, the observed extended absorption profile may indicate that the long dark gaps in \citet{zhu_long_2022} are similar to simulated dark gaps that are created by neutral islands.

\begin{figure}
    \centering
    \includegraphics[width=3.2in]{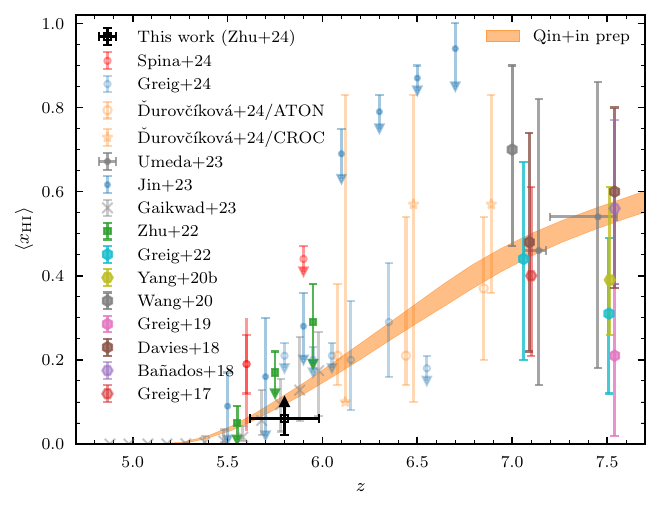}
    \vspace{-0.5cm}
    \caption{Lower limit constraint on the volume-averaged neutral hydrogen fraction from the damping wing-like profile in the stacked \lya\ forest based on \lyb\ dark gaps over $5.6<z<6.0$. We also plot $\langle x_{\rm HI} \rangle$ constraints from recent observations for reference: upper limit constraints from \lya\ + \lyb\ dark gap statistics \citep{zhu_long_2022}, upper limit constraints from the fraction of \lya\ + \lyb\ dark pixels \citep{jin_nearly_2023},  measurements from high-$z$ quasar damping wings \citep{greig_are_2017,greig_constraints_2019,greig_igm_2022,banados_800-million-solar-mass_2018,davies_quantitative_2018,wang_significantly_2020,yang_poniuaena_2020,durovcikova_chronicling_2024-1,greig_igm_2024}, constraints based on \lya\ opacities\citep{gaikwad_measuring_2023}, and measurements from the \lya\ damping wing of high-$z$ galaxies \citep{umeda_jwst_2023}. We also include the measurements based on the stacked \lya\ forest using a different method in \citet{spina_damping_2024}. 
    The orange shaded region plot the posterior from inference based on multiple reionization observables (Y.~Qin et al.~in prep.). 
    Data points in this figure are slightly displaced along $z$ for display purpose.
    }
    \label{fig:xHI}
\end{figure}

\section{Lower Limit Constraint on Neutral Hydrogen Fraction}\label{sec:lower_limit}

We can use the observed damping wing-like profile to constrain the hydrogen neutral fraction in unbiased locations within the IGM.  
To do this, we generate dark gaps following the observed gap length distribution for different ranges of $x_{\rm HI, gap}$, specifically $0.1 \leq x_{\rm HI, gap} < 0.2$, $0.2 \leq x_{\rm HI, gap} < 0.3$, ..., up to $0.9 \leq x_{\rm HI, gap} < 1.0$, and create mock \lya\ forest stacks for each range using the RT simulation used for our first model. 
We find that only $x_{\rm HI, gap}$ values greater than 0.4 can predict $v_{\rm ext}$ consistent with the observed value within the 95\% limit. After convolving $x_{\rm HI, gap}\ge 0.4$ with the 
fraction of the \lya\ forest subtended by dark gaps longer than the mean length of dark gaps used in this work, $f_{L\geq 15\cmpch}=0.152 \pm 0.098$, \citep{zhu_long_2022}
we derive a lower limit on the volume-averaged neutral hydrogen fraction of $\langle x_{\rm HI} \rangle \ge 0.061 \pm 0.039$ at $z=5.8$. This constraint is conservative in that it excludes potential neutral hydrogen in dark gaps with $L<7 \cmpch$. We also emphasize that the simulations we are using include significant UVB fluctuations, which helps us to account for the possibility that a  fraction of the dark gaps will be produced by ionized gas.

Figure \ref{fig:xHI} displays our constraint alongside other recent $\langle x_{\rm HI} \rangle$ measurements. Our constraint on $\langle x_{\rm HI} \rangle$ suggests that the IGM must be at least 6.1\% neutral even down to $z=5.8$. Alongside upper limit constraints from dark gaps and dark pixels \citep{zhu_long_2022,jin_nearly_2023}, these observations accommodate a wide range of late-ending reionization histories, with $\langle x_{\rm HI} \rangle$ constraints ranging from approximately 6\% to 20\% near $z=5.8$.
We also note that the inference based on multiple recent reionization observables (Y.~Qin et al.~in prep.), which includes the mean transmission in the \lya\ forest, galaxy UV luminosity functions, and CMB observations, aligns well with these $\langle x_{\rm HI} \rangle$ constraints.

\section{Summary}\label{sec:summary_p4}

We have identified a damping wing-like profile in the stacked \lya\ forest associated with \lyb\ dark gaps at $z<6$, using data from the XQR-30 program and archival spectra. A comparison with mock spectra generated from simulations suggests that this damping wing-like profile most likely arises from dark gaps containing neutral hydrogen islands, as opposed to fully ionized regions whose high opacities are the result of high densities and/or a locally low UVB. We further ascertain that such a profile necessitates a neutral hydrogen fraction of at least $6.1 \pm 3.9\%$ at $z=5.8$. Combined with the IGM damping wing discovered toward an individual quasar sightline at $z=5.878$ (see the companion paper, \citealp{becker_damping_2024}), our results provide robust evidence that reionization extends to $z<6$. 

\section*{Acknowledgements}
We thank the anonymous reviewer for their helpful comments.
Y.Z.~and G.D.B.~were supported by the National Science Foundation through grant AST-1751404. Y.Z.~was also supported by the NSF through award SOSPADA-029 from the NRAO. 
G.K.~and M.G.H.~have been partially supported by the DAE-STFC project ``Building Indo-UK collaborations towards the Square Kilometre Array'' (STFC grant reference ST/Y004191/1).
Parts of this research were supported by the Australian Research Council Centre of Excellence for All Sky Astrophysics in 3 Dimensions (ASTRO 3D), through project no. CE170100013. 
B.S.~and S.E.I.B.~are supported by the Deutsche Forschungsge-meinschaft (DFG) under Emmy Noether grant number BO 5771/1-1.
F.W.~acknowledges support from NSF Grant AST-2308258. 
For the purpose of open access, the authors have applied a Creative Commons Attribution (CC BY) licence to any Author Accepted Manuscript version arising from this submission.

This work is based on observations collected at the European Southern Observatory under ESO programme 1103.A-0817.
Some of the data presented herein were obtained at Keck Observatory, which is a private 501(c)3 non-profit organization operated as a scientific partnership among the California Institute of Technology, the University of California, and the National Aeronautics and Space Administration. The Observatory was made possible by the generous financial support of the W. M. Keck Foundation. The authors wish to recognize and acknowledge the very significant cultural role and reverence that the summit of Maunakea has always had within the Native Hawaiian community. We are most fortunate to have the opportunity to conduct observations from this mountain. This research has made use of the Keck Observatory Archive (KOA), which is operated by the W. M. Keck Observatory and the NASA Exoplanet Science Institute (NExScI), under contract with the National Aeronautics and Space Administration.

We respectfully acknowledge the University of Arizona is on the land and territories of Indigenous peoples. Today, Arizona is home to 22 federally recognized tribes, with Tucson being home to the O'odham and the Yaqui. Committed to diversity and inclusion, the University strives to build sustainable relationships with sovereign Native Nations and Indigenous communities through education offerings, partnerships, and community service.

The simulations used in this work were performed using the Joliot Curie supercomputer at the Tr{\`e}s Grand Centre de Calcul (TGCC) and the Cambridge Service for Data Driven Discovery (CSD3), part of which is operated by the University of Cambridge Research Computing on behalf of the STFC DiRAC HPC Facility (www.dirac.ac.uk).  We acknowledge the Partnership for Advanced Computing in Europe (PRACE) for awarding us time on Joliot Curie in the 16th call. The DiRAC component of CSD3 was funded by BEIS capital funding via STFC capital grants ST/P002307/1 and ST/R002452/1 and STFC operations grant ST/R00689X/1.  This work also used the DiRAC@Durham facility managed by the Institute for Computational Cosmology on behalf of the STFC DiRAC HPC Facility. The equipment was funded by BEIS capital funding via STFC capital grants ST/P002293/1 and ST/R002371/1, Durham University and STFC operations grant ST/R000832/1. DiRAC is part of the National e-Infrastructure.

\textit{Software}: Astropy \citep{astropy_collaboration_astropy_2013}, Matplotlib \citep{hunter_matplotlib_2007}, NumPy \citep{van_der_walt_numpy_2011}, SpectRes \citep{carnall_spectres_2017}.

\section*{Data Availability}

The raw data used in this work are available from the ESO archive at \url{http://archive.eso.org} and KOA at \url{https://koa.ipac.caltech.edu}. The reduced X-Shooter spectra are available through a public GitHub repository of the XQR-30 spectra at \url{https://github.com/XQR-30/Spectra}. The reduced Keck spectra can be obtained from the corresponding author upon reasonable request.




\begin{thebibliography}{}
\makeatletter
\relax
\def\mn@urlcharsother{\let\do\@makeother \do\$\do\&\do\#\do\^\do\_\do\%\do\~}
\def\mn@doi{\begingroup\mn@urlcharsother \@ifnextchar [ {\mn@doi@} {\mn@doi@[]}}
\def\mn@doi@[#1]#2{\def\@tempa{#1}\ifx\@tempa\@empty \href {http://dx.doi.org/#2} {doi:#2}\else \href {http://dx.doi.org/#2} {#1}\fi \endgroup}
\def\mn@eprint#1#2{\mn@eprint@#1:#2::\@nil}
\def\mn@eprint@arXiv#1{\href {http://arxiv.org/abs/#1} {{\tt arXiv:#1}}}
\def\mn@eprint@dblp#1{\href {http://dblp.uni-trier.de/rec/bibtex/#1.xml} {dblp:#1}}
\def\mn@eprint@#1:#2:#3:#4\@nil{\def\@tempa {#1}\def\@tempb {#2}\def\@tempc {#3}\ifx \@tempc \@empty \let \@tempc \@tempb \let \@tempb \@tempa \fi \ifx \@tempb \@empty \def\@tempb {arXiv}\fi \@ifundefined {mn@eprint@\@tempb}{\@tempb:\@tempc}{\expandafter \expandafter \csname mn@eprint@\@tempb\endcsname \expandafter{\@tempc}}}

\bibitem[\protect\citeauthoryear{Asthana, Haehnelt, Kulkarni, Aubert, Bolton  \& Keating}{Asthana et~al.}{2024}]{asthana_late-end_2024}
Asthana S.,  Haehnelt M.~G.,  Kulkarni G.,  Aubert D.,  Bolton J.~S.,   Keating L.~C.,  2024, Late-End Reionization with {{ATON-HE}}: Towards Constraints from {{Lyman-}}\${\textbackslash}alpha\$ Emitters Observed with {{JWST}}, \mn@doi{10.48550/arXiv.2404.06548}

\bibitem[\protect\citeauthoryear{{Astropy Collaboration} et~al.,}{{Astropy Collaboration} et~al.}{2013}]{astropy_collaboration_astropy_2013}
{Astropy Collaboration} et~al., 2013, \mn@doi [\aap] {10.1051/0004-6361/201322068}, 558, A33

\bibitem[\protect\citeauthoryear{Ba{\~n}ados et~al.,}{Ba{\~n}ados et~al.}{2018}]{banados_800-million-solar-mass_2018}
Ba{\~n}ados E.,  et~al., 2018, \mn@doi [\nat] {10.1038/nature25180}, 553, 473

\bibitem[\protect\citeauthoryear{Becker, Bolton, Madau, Pettini, {Ryan-Weber}  \& Venemans}{Becker et~al.}{2015}]{becker_evidence_2015}
Becker G.~D.,  Bolton J.~S.,  Madau P.,  Pettini M.,  {Ryan-Weber} E.~V.,   Venemans B.~P.,  2015, \mn@doi [\mnras] {10.1093/mnras/stu2646}, 447, 3402

\bibitem[\protect\citeauthoryear{Becker, Davies, Furlanetto, Malkan, Boera  \& Douglass}{Becker et~al.}{2018}]{becker_evidence_2018}
Becker G.~D.,  Davies F.~B.,  Furlanetto S.~R.,  Malkan M.~A.,  Boera E.,   Douglass C.,  2018, \mn@doi [\apj] {10.3847/1538-4357/aacc73}, 863, 92

\bibitem[\protect\citeauthoryear{Becker et~al.,}{Becker et~al.}{2019}]{becker_evolution_2019}
Becker G.~D.,  et~al., 2019, \mn@doi [\apj] {10.3847/1538-4357/ab3eb5}, 883, 163

\bibitem[\protect\citeauthoryear{Becker, D'Aloisio, Christenson, Zhu, Worseck  \& Bolton}{Becker et~al.}{2021}]{becker_mean_2021}
Becker G.~D.,  D'Aloisio A.,  Christenson H.~M.,  Zhu Y.,  Worseck G.,   Bolton J.~S.,  2021, \mn@doi [\mnras] {10.1093/mnras/stab2696}, 508, 1853

\bibitem[\protect\citeauthoryear{Becker, Bolton, Zhu  \& Hashemi}{Becker et~al.}{2024}]{becker_damping_2024}
Becker G.~D.,  Bolton J.~S.,  Zhu Y.,   Hashemi S.,  2024, Damping Wing Absorption Associated with a Giant {{Ly}}\${\textbackslash}alpha\$ Trough at \$z {$<$} 6\$: Direct Evidence for Late-Ending Reionization, \mn@doi{10.48550/arXiv.2405.08885}

\bibitem[\protect\citeauthoryear{Bosman}{Bosman}{2021}]{bosman_constraints_2021-1}
Bosman S. E.~I.,  2021, Constraints on the Mean Free Path of Ionising Photons at \$z{\textbackslash}sim6\$ Using Limits on Individual Free Paths, \mn@doi{10.48550/arXiv.2108.12446}

\bibitem[\protect\citeauthoryear{Bosman, Fan, Jiang, Reed, Matsuoka, Becker  \& Haehnelt}{Bosman et~al.}{2018}]{bosman_new_2018}
Bosman S. E.~I.,  Fan X.,  Jiang L.,  Reed S.,  Matsuoka Y.,  Becker G.,   Haehnelt M.,  2018, \mn@doi [\mnras] {10.1093/mnras/sty1344}, 479, 1055

\bibitem[\protect\citeauthoryear{Bosman, {\v D}urov{\v c}{\'i}kov{\'a}, Davies  \& Eilers}{Bosman et~al.}{2021}]{bosman_comparison_2021}
Bosman S. E.~I.,  {\v D}urov{\v c}{\'i}kov{\'a} D.,  Davies F.~B.,   Eilers A.~C.,  2021, \mn@doi [\mnras] {10.1093/mnras/stab572}, 503, 2077

\bibitem[\protect\citeauthoryear{Bosman et~al.,}{Bosman et~al.}{2022}]{bosman_hydrogen_2022}
Bosman S. E.~I.,  et~al., 2022, \mn@doi [\mnras] {10.1093/mnras/stac1046}, 514, 55

\bibitem[\protect\citeauthoryear{Bouwens, Illingworth, Oesch, Caruana, Holwerda, Smit  \& Wilkins}{Bouwens et~al.}{2015}]{bouwens_reionization_2015}
Bouwens R.~J.,  Illingworth G.~D.,  Oesch P.~A.,  Caruana J.,  Holwerda B.,  Smit R.,   Wilkins S.,  2015, \mn@doi [\apj] {10.1088/0004-637X/811/2/140}, 811, 140

\bibitem[\protect\citeauthoryear{Cain, D'Aloisio, Gangolli  \& Becker}{Cain et~al.}{2021}]{cain_short_2021}
Cain C.,  D'Aloisio A.,  Gangolli N.,   Becker G.~D.,  2021, \mn@doi [\apj] {10.3847/2041-8213/ac1ace}, 917, L37

\bibitem[\protect\citeauthoryear{Cain, D'Aloisio, Lopez, Gangolli  \& Roth}{Cain et~al.}{2023a}]{cain_rise_2023-1}
Cain C.,  D'Aloisio A.,  Lopez G.,  Gangolli N.,   Roth J.~T.,  2023a, On the Rise and Fall of Galactic Ionizing Output at the End of Reionization, \mn@doi{10.48550/arXiv.2311.13638}

\bibitem[\protect\citeauthoryear{Cain, D'Aloisio, Gangolli  \& McQuinn}{Cain et~al.}{2023b}]{cain_morphology_2023}
Cain C.,  D'Aloisio A.,  Gangolli N.,   McQuinn M.,  2023b, \mn@doi [\mnras] {10.1093/mnras/stad1057}, 522, 2047

\bibitem[\protect\citeauthoryear{Carnall}{Carnall}{2017}]{carnall_spectres_2017}
Carnall A.~C.,  2017, arXiv e-prints, p. arXiv:1705.05165

\bibitem[\protect\citeauthoryear{Choudhury, Paranjape  \& Bosman}{Choudhury et~al.}{2021}]{choudhury_studying_2021}
Choudhury T.~R.,  Paranjape A.,   Bosman S. E.~I.,  2021, \mn@doi [\mnras] {10.1093/mnras/stab045}, 501, 5782

\bibitem[\protect\citeauthoryear{Christenson, Becker, Furlanetto, Davies, Malkan, Zhu, Boera  \& Trapp}{Christenson et~al.}{2021}]{christenson_constraints_2021}
Christenson H.~M.,  Becker G.~D.,  Furlanetto S.~R.,  Davies F.~B.,  Malkan M.~A.,  Zhu Y.,  Boera E.,   Trapp A.,  2021, \mn@doi [\apj] {10.3847/1538-4357/ac2a34}, 923, 87

\bibitem[\protect\citeauthoryear{Christenson et~al.,}{Christenson et~al.}{2023}]{christenson_relationship_2023-1}
Christenson H.~M.,  et~al., 2023, \mn@doi [\apj] {10.3847/1538-4357/acf450}, 955, 138

\bibitem[\protect\citeauthoryear{Cooper, Simcoe, Cooksey, Bordoloi, Miller, Furesz, Turner  \& Ba{\~n}ados}{Cooper et~al.}{2019}]{cooper_heavy_2019}
Cooper T.~J.,  Simcoe R.~A.,  Cooksey K.~L.,  Bordoloi R.,  Miller D.~R.,  Furesz G.,  Turner M.~L.,   Ba{\~n}ados E.,  2019, \mn@doi [\apj] {10.3847/1538-4357/ab3402}, 882, 77

\bibitem[\protect\citeauthoryear{D'Odorico et~al.,}{D'Odorico et~al.}{2023}]{dodorico_xqr-30_2023}
D'Odorico V.,  et~al., 2023, \mn@doi [\mnras] {10.1093/mnras/stad1468}, 523, 1399

\bibitem[\protect\citeauthoryear{Davies et~al.,}{Davies et~al.}{2018}]{davies_quantitative_2018}
Davies F.~B.,  et~al., 2018, \mn@doi [\apj] {10.3847/1538-4357/aad6dc}, 864, 142

\bibitem[\protect\citeauthoryear{Davies et~al.,}{Davies et~al.}{2023a}]{davies_xqr-30_2023}
Davies R.~L.,  et~al., 2023a, \mn@doi [\mnras] {10.1093/mnras/stac3662}, 521, 289

\bibitem[\protect\citeauthoryear{Davies et~al.,}{Davies et~al.}{2023b}]{davies_examining_2023-1}
Davies R.~L.,  et~al., 2023b, \mn@doi [\mnras] {10.1093/mnras/stad294}, 521, 314

\bibitem[\protect\citeauthoryear{Davies et~al.,}{Davies et~al.}{2024}]{davies_constraints_2024}
Davies F.~B.,  et~al., 2024, \mn@doi [\apj] {10.3847/1538-4357/ad1d5d}, 965, 134

\bibitem[\protect\citeauthoryear{Decarli et~al.,}{Decarli et~al.}{2018}]{decarli_alma_2018}
Decarli R.,  et~al., 2018, \mn@doi [\apj] {10.3847/1538-4357/aaa5aa}, 854, 97

\bibitem[\protect\citeauthoryear{{\v D}urov{\v c}{\'i}kov{\'a} et~al.,}{{\v D}urov{\v c}{\'i}kov{\'a} et~al.}{2024}]{durovcikova_chronicling_2024-1}
{\v D}urov{\v c}{\'i}kov{\'a} D.,  et~al., 2024, Chronicling the Reionization History at \$6{\textbackslash}lesssim z {\textbackslash}lesssim 7\$ with Emergent Quasar Damping Wings, \mn@doi{10.48550/arXiv.2401.10328}

\bibitem[\protect\citeauthoryear{Eilers, Davies  \& Hennawi}{Eilers et~al.}{2018}]{eilers_opacity_2018}
Eilers A.-C.,  Davies F.~B.,   Hennawi J.~F.,  2018, \mn@doi [\apj] {10.3847/1538-4357/aad4fd}, 864, 53

\bibitem[\protect\citeauthoryear{Eilers et~al.,}{Eilers et~al.}{2020}]{eilers_detecting_2020}
Eilers A.-C.,  et~al., 2020, \mn@doi [\apj] {10.3847/1538-4357/aba52e}, 900, 37

\bibitem[\protect\citeauthoryear{Fan et~al.,}{Fan et~al.}{2006}]{fan_constraining_2006}
Fan X.,  et~al., 2006, \mn@doi [\aj] {10.1086/504836}, 132, 117

\bibitem[\protect\citeauthoryear{Fan, Ba{\~n}ados  \& Simcoe}{Fan et~al.}{2023}]{fan_quasars_2023}
Fan X.,  Ba{\~n}ados E.,   Simcoe R.~A.,  2023, \mn@doi [\araa] {10.1146/annurev-astro-052920-102455}, 61, 373

\bibitem[\protect\citeauthoryear{Finkelstein et~al.,}{Finkelstein et~al.}{2019}]{finkelstein_conditions_2019}
Finkelstein S.~L.,  et~al., 2019, \mn@doi [\apj] {10.3847/1538-4357/ab1ea8}, 879, 36

\bibitem[\protect\citeauthoryear{Gaikwad et~al.,}{Gaikwad et~al.}{2023}]{gaikwad_measuring_2023}
Gaikwad P.,  et~al., 2023, \mn@doi [\mnras] {10.1093/mnras/stad2566}, 525, 4093

\bibitem[\protect\citeauthoryear{Gnedin}{Gnedin}{2022}]{gnedin_cosmic_2022-1}
Gnedin N.~Y.,  2022, \mn@doi [\apj] {10.3847/1538-4357/ac8d0a}, 937, 17

\bibitem[\protect\citeauthoryear{Greig, Mesinger, Haiman  \& Simcoe}{Greig et~al.}{2017}]{greig_are_2017}
Greig B.,  Mesinger A.,  Haiman Z.,   Simcoe R.~A.,  2017, \mn@doi [\mnras] {10.1093/mnras/stw3351}, 466, 4239

\bibitem[\protect\citeauthoryear{Greig, Mesinger  \& Ba{\~n}ados}{Greig et~al.}{2019}]{greig_constraints_2019}
Greig B.,  Mesinger A.,   Ba{\~n}ados E.,  2019, \mn@doi [\mnras] {10.1093/mnras/stz230}, 484, 5094

\bibitem[\protect\citeauthoryear{Greig, Mesinger, Davies, Wang, Yang  \& Hennawi}{Greig et~al.}{2022}]{greig_igm_2022}
Greig B.,  Mesinger A.,  Davies F.~B.,  Wang F.,  Yang J.,   Hennawi J.~F.,  2022, \mn@doi [\mnras] {10.1093/mnras/stac825}, 512, 5390

\bibitem[\protect\citeauthoryear{Greig et~al.,}{Greig et~al.}{2024}]{greig_igm_2024}
Greig B.,  et~al., 2024, \mn@doi [\mnras] {10.1093/mnras/stae1080}, 530, 3208

\bibitem[\protect\citeauthoryear{Hunter}{Hunter}{2007}]{hunter_matplotlib_2007}
Hunter J.~D.,  2007, \mn@doi [CSE] {10.1109/MCSE.2007.55}, 9, 90

\bibitem[\protect\citeauthoryear{Jin et~al.,}{Jin et~al.}{2023}]{jin_nearly_2023}
Jin X.,  et~al., 2023, \mn@doi [\apj] {10.3847/1538-4357/aca678}, 942, 59

\bibitem[\protect\citeauthoryear{Kashino, Lilly, Shibuya, Ouchi  \& Kashikawa}{Kashino et~al.}{2020}]{kashino_evidence_2020}
Kashino D.,  Lilly S.~J.,  Shibuya T.,  Ouchi M.,   Kashikawa N.,  2020, \mn@doi [\apj] {10.3847/1538-4357/ab5a7d}, 888, 6

\bibitem[\protect\citeauthoryear{Keating, Weinberger, Kulkarni, Haehnelt, Chardin  \& Aubert}{Keating et~al.}{2020}]{keating_long_2020}
Keating L.~C.,  Weinberger L.~H.,  Kulkarni G.,  Haehnelt M.~G.,  Chardin J.,   Aubert D.,  2020, \mn@doi [\mnras] {10.1093/mnras/stz3083}, 491, 1736

\bibitem[\protect\citeauthoryear{Kulkarni, Keating, Haehnelt, Bosman, Puchwein, Chardin  \& Aubert}{Kulkarni et~al.}{2019}]{kulkarni_large_2019}
Kulkarni G.,  Keating L.~C.,  Haehnelt M.~G.,  Bosman S. E.~I.,  Puchwein E.,  Chardin J.,   Aubert D.,  2019, \mn@doi [\mnras] {10.1093/mnrasl/slz025}, 485, L24

\bibitem[\protect\citeauthoryear{Malloy \& Lidz}{Malloy \& Lidz}{2015}]{malloy_how_2015}
Malloy M.,  Lidz A.,  2015, \mn@doi [\apj] {10.1088/0004-637X/799/2/179}, 799, 179

\bibitem[\protect\citeauthoryear{McGreer et~al.,}{McGreer et~al.}{2013}]{mcgreer_z_2013}
McGreer I.~D.,  et~al., 2013, \mn@doi [\apj] {10.1088/0004-637X/768/2/105}, 768, 105

\bibitem[\protect\citeauthoryear{{Miralda-Escud{\'e}}}{{Miralda-Escud{\'e}}}{1998}]{miralda-escude_reionization_1998}
{Miralda-Escud{\'e}} J.,  1998, \mn@doi [\apj] {10.1086/305799}, 501, 15

\bibitem[\protect\citeauthoryear{{Miralda-Escud{\'e}} \& Rees}{{Miralda-Escud{\'e}} \& Rees}{1998}]{miralda-escude_searching_1998}
{Miralda-Escud{\'e}} J.,  Rees M.~J.,  1998, \mn@doi [\apj] {10.1086/305458}, 497, 21

\bibitem[\protect\citeauthoryear{Mu{\~n}oz, Mirocha, Chisholm, Furlanetto  \& Mason}{Mu{\~n}oz et~al.}{2024}]{munoz_reionization_2024-1}
Mu{\~n}oz J.~B.,  Mirocha J.,  Chisholm J.,  Furlanetto S.~R.,   Mason C.,  2024, Reionization after {{JWST}}: A Photon Budget Crisis?, \mn@doi{10.48550/arXiv.2404.07250}

\bibitem[\protect\citeauthoryear{Nasir \& D'Aloisio}{Nasir \& D'Aloisio}{2020}]{nasir_observing_2020}
Nasir F.,  D'Aloisio A.,  2020, \mn@doi [\mnras] {10.1093/mnras/staa894}, 494, 3080

\bibitem[\protect\citeauthoryear{{Planck Collaboration} et~al.,}{{Planck Collaboration} et~al.}{2014}]{planck_collaboration_planck_2014}
{Planck Collaboration} et~al., 2014, \mn@doi [\aap] {10.1051/0004-6361/201321591}, 571, A16

\bibitem[\protect\citeauthoryear{{Planck Collaboration} et~al.,}{{Planck Collaboration} et~al.}{2020}]{planck_collaboration_planck_2020}
{Planck Collaboration} et~al., 2020, \mn@doi [\aap] {10.1051/0004-6361/201833910}, 641, A6

\bibitem[\protect\citeauthoryear{Prochaska, Gawiser, Wolfe, Cooke  \& Gelino}{Prochaska et~al.}{2003}]{prochaska_esikeck_2003}
Prochaska J.~X.,  Gawiser E.,  Wolfe A.~M.,  Cooke J.,   Gelino D.,  2003, \mn@doi [\apjs] {10.1086/375839}, 147, 227

\bibitem[\protect\citeauthoryear{Puchwein et~al.,}{Puchwein et~al.}{2023}]{puchwein_sherwood-relics_2023}
Puchwein E.,  et~al., 2023, \mn@doi [\mnras] {10.1093/mnras/stac3761}, 519, 6162

\bibitem[\protect\citeauthoryear{Qin, Mesinger, Bosman  \& Viel}{Qin et~al.}{2021}]{qin_reionization_2021}
Qin Y.,  Mesinger A.,  Bosman S. E.~I.,   Viel M.,  2021, \mn@doi [\mnras] {10.1093/mnras/stab1833}, 506, 2390

\bibitem[\protect\citeauthoryear{Robertson, Ellis, Furlanetto  \& Dunlop}{Robertson et~al.}{2015}]{robertson_cosmic_2015}
Robertson B.~E.,  Ellis R.~S.,  Furlanetto S.~R.,   Dunlop J.~S.,  2015, \mn@doi [\apj] {10.1088/2041-8205/802/2/L19}, 802, L19

\bibitem[\protect\citeauthoryear{Roth, D'Aloisio, Cain, Wilson, Zhu  \& Becker}{Roth et~al.}{2024}]{roth_effect_2024}
Roth J.~T.,  D'Aloisio A.,  Cain C.,  Wilson B.,  Zhu Y.,   Becker G.~D.,  2024, \mn@doi [\mnras] {10.1093/mnras/stae1194}, 530, 5209

\bibitem[\protect\citeauthoryear{Satyavolu, Kulkarni, Keating  \& Haehnelt}{Satyavolu et~al.}{2023}]{satyavolu_robustness_2023}
Satyavolu S.,  Kulkarni G.,  Keating L.~C.,   Haehnelt M.~G.,  2023, Robustness of Direct Measurements of the Mean Free Path of Ionizing Photons in the Epoch of Reionization (\mn@eprint {arxiv} {2311.06344})

\bibitem[\protect\citeauthoryear{Sebastian et~al.,}{Sebastian et~al.}{2024}]{sebastian_e-xqr-30_2024}
Sebastian A.~M.,  et~al., 2024, \mn@doi [\mnras] {10.1093/mnras/stae789}, 530, 1829

\bibitem[\protect\citeauthoryear{Sheinis, Bolte, Epps, Kibrick, Miller, Radovan, Bigelow  \& Sutin}{Sheinis et~al.}{2002}]{sheinis_esi_2002}
Sheinis A.~I.,  Bolte M.,  Epps H.~W.,  Kibrick R.~I.,  Miller J.~S.,  Radovan M.~V.,  Bigelow B.~C.,   Sutin B.~M.,  2002, \mn@doi [\pasp] {10.1086/341706}, 114, 851

\bibitem[\protect\citeauthoryear{Shen et~al.,}{Shen et~al.}{2019}]{shen_gemini_2019}
Shen Y.,  et~al., 2019, \mn@doi [\apj] {10.3847/1538-4357/ab03d9}, 873, 35

\bibitem[\protect\citeauthoryear{Spina, Bosman, Davies, Gaikwad  \& Zhu}{Spina et~al.}{2024}]{spina_damping_2024}
Spina B.,  Bosman S. E.~I.,  Davies F.~B.,  Gaikwad P.,   Zhu Y.,  2024, Damping Wings in the {{Lyman-}}\{{\textbackslash}alpha\} Forest: A Model-Independent Measurement of the Neutral Fraction at 5.4, \mn@doi{10.48550/arXiv.2405.12273}

\bibitem[\protect\citeauthoryear{Stark}{Stark}{2016}]{stark_galaxies_2016}
Stark D.~P.,  2016, \mn@doi [\araa] {10.1146/annurev-astro-081915-023417}, 54, 761

\bibitem[\protect\citeauthoryear{Umeda, Ouchi, Nakajima, Harikane, Ono, Xu, Isobe  \& Zhang}{Umeda et~al.}{2023}]{umeda_jwst_2023}
Umeda H.,  Ouchi M.,  Nakajima K.,  Harikane Y.,  Ono Y.,  Xu Y.,  Isobe Y.,   Zhang Y.,  2023, {{JWST Measurements}} of {{Neutral Hydrogen Fractions}} and {{Ionized Bubble Sizes}} at \$z=7-12\$ {{Obtained}} with {{Ly}}\${\textbackslash}alpha\$ {{Damping Wing Absorptions}} in 26 {{Bright Continuum Galaxies}} (\mn@eprint {arxiv} {2306.00487})

\bibitem[\protect\citeauthoryear{Vernet et~al.,}{Vernet et~al.}{2011}]{vernet_x-shooter_2011}
Vernet J.,  et~al., 2011, \mn@doi [\aap] {10.1051/0004-6361/201117752}, 536, A105

\bibitem[\protect\citeauthoryear{Wang et~al.,}{Wang et~al.}{2010}]{wang_molecular_2010}
Wang R.,  et~al., 2010, \mn@doi [\apj] {10.1088/0004-637X/714/1/699}, 714, 699

\bibitem[\protect\citeauthoryear{Wang et~al.,}{Wang et~al.}{2013}]{wang_star_2013}
Wang R.,  et~al., 2013, \mn@doi [\apj] {10.1088/0004-637X/773/1/44}, 773, 44

\bibitem[\protect\citeauthoryear{Wang et~al.,}{Wang et~al.}{2020}]{wang_significantly_2020}
Wang F.,  et~al., 2020, \mn@doi [\apj] {10.3847/1538-4357/ab8c45}, 896, 23

\bibitem[\protect\citeauthoryear{Weinberger, Haehnelt  \& Kulkarni}{Weinberger et~al.}{2019}]{weinberger_modelling_2019}
Weinberger L.~H.,  Haehnelt M.~G.,   Kulkarni G.,  2019, \mn@doi [\mnras] {10.1093/mnras/stz481}, 485, 1350

\bibitem[\protect\citeauthoryear{Yang et~al.,}{Yang et~al.}{2020a}]{yang_poniuaena_2020}
Yang J.,  et~al., 2020a, \mn@doi [\apjl] {10.3847/2041-8213/ab9c26}, 897, L14

\bibitem[\protect\citeauthoryear{Yang et~al.,}{Yang et~al.}{2020b}]{yang_measurements_2020}
Yang J.,  et~al., 2020b, \mn@doi [\apj] {10.3847/1538-4357/abbc1b}, 904, 26

\bibitem[\protect\citeauthoryear{Zhu et~al.,}{Zhu et~al.}{2021}]{zhu_chasing_2021}
Zhu Y.,  et~al., 2021, \mn@doi [\apj] {10.3847/1538-4357/ac26c2}, 923, 223

\bibitem[\protect\citeauthoryear{Zhu et~al.,}{Zhu et~al.}{2022}]{zhu_long_2022}
Zhu Y.,  et~al., 2022, \mn@doi [\apj] {10.3847/1538-4357/ac6e60}, 932, 76

\bibitem[\protect\citeauthoryear{Zhu et~al.,}{Zhu et~al.}{2023}]{zhu_probing_2023}
Zhu Y.,  et~al., 2023, \mn@doi [\apj] {10.3847/1538-4357/aceef4}, 955, 115

\bibitem[\protect\citeauthoryear{{de Belsunce}, Gratton, Coulton  \& Efstathiou}{{de Belsunce} et~al.}{2021}]{de_belsunce_inference_2021}
{de Belsunce} R.,  Gratton S.,  Coulton W.,   Efstathiou G.,  2021, \mn@doi [\mnras] {10.1093/mnras/stab2215}, 507, 1072

\bibitem[\protect\citeauthoryear{{van der Walt}, Colbert  \& Varoquaux}{{van der Walt} et~al.}{2011}]{van_der_walt_numpy_2011}
{van der Walt} S.,  Colbert S.~C.,   Varoquaux G.,  2011, \mn@doi [CSE] {10.1109/MCSE.2011.37}, 13, 22

\makeatother
\end{thebibliography}
\input{output.bbl}








\bsp	
\label{lastpage}
\end{document}